\documentclass{sigchi-ext}
\usepackage[T1]{fontenc}
\usepackage{textcomp}
\usepackage[scaled=.92]{helvet} % for proper fonts
\usepackage{graphicx} % for EPS use the graphics package instead
\usepackage{balance}  % for useful for balancing the last columns
\usepackage{booktabs} % for pretty table rules
\usepackage{ccicons}  % for Creative Commons citation icons
\usepackage{ragged2e} % for tighter hyphenation

\def\plaintitle{Large Language Models Cannot Explain Themselves} 
\def\emptyauthor{}
\def\plainkeywords{exoplanations; mechanismal explanations; co-audit; AI safety; explainable AI; XAI; critical thinking}

\title{Large Language Models Cannot Explain Themselves}

\numberofauthors{6}
\author{%
  \alignauthor{%
    \textbf{Advait Sarkar}\\
    \affaddr{Microsoft Research} \\
    \affaddr{Cambridge, United Kingdom} \\~\\
    \affaddr{University of Cambridge} \\
    \affaddr{Cambridge, United Kingdom} \\~\\
    \affaddr{University College London} \\
    \affaddr{London, United Kingdom} \\~\\
    \email{advait@microsoft.com} 
    }
}

\definecolor{linkColor}{RGB}{6,125,233}
\hypersetup{%
  pdftitle={\plaintitle},
  pdfauthor={\emptyauthor},
  pdfkeywords={\plainkeywords},
  bookmarksnumbered,
  pdfstartview={FitH},
  colorlinks,
  citecolor=black,
  filecolor=black,
  linkcolor=black,
  urlcolor=linkColor,
  breaklinks=true,
}

\usepackage{todonotes}
\usepackage{xurl}

\emergencystretch=\maxdimen
\begin{document}

%% For the camera ready, use the commands provided by the ACM in the Permission Release Form.
\CopyrightYear{2024}
\setcopyright{rightsretained}
\conferenceinfo{HCXAI workshop at CHI'24}{Honolulu, HI, USA}
% \isbn{}
% \doi{}
%% Then override the default copyright message with the \acmcopyright command.
\copyrightinfo{\acmcopyright}

\maketitle

% Uncomment to disable hyphenation (not recommended)
% https://twitter.com/anjirokhan/status/546046683331973120
\RaggedRight{} 
% \justifying{}

% Do not change the page size or page settings.
\begin{abstract}
Large language models can be prompted to produce text. They can also be prompted to produce ``explanations'' of their output. But these are not really explanations, because they do not accurately reflect the mechanical process underlying the prediction. The illusion that they reflect the reasoning process can result in significant harms. These ``explanations'' can be valuable, but for promoting critical thinking rather than for understanding the model. I propose a recontextualisation of these ``explanations'', using the term ``exoplanations'' to draw attention to their exogenous nature. I discuss some implications for design and technology, such as the inclusion of appropriate guardrails and responses when models are prompted to generate explanations.
\end{abstract}


% ACM Classfication



% Print the classficiation codes
% Please use the 2012 Classifiers and see this link to embed them in the text: \url{https://dl.acm.org/ccs/ccs_flat.cfm}

\section{The illusion of explanation}

In the context of Artificial Intelligence (AI), the term ``explanation'' can encompass many types of information. The most well-studied category of explanations is concerned with providing descriptions of the structure of a model, its training data, and most commonly, elaboration of any given output in terms of the algorithmic process followed to produce that output \cite{sarkar2022explainable}. What these explanations have in common is that they aim to faithfully represent some aspect of the real underlying algorithmic mechanism of an AI model. Let us therefore refer to these as \emph{mechanismal explanations}.\footnote{The aim of this awkward construction is to emphasise that they are explanations \emph{of} a mechanism; using the term ``mechanical'' or ``mechanistic'' might connote that the explanations themselves are generated mechanically, which is usually true but not relevant.}

\marginpar{%
\RaggedRight{} 
  \vspace{-140pt} \fbox{%
    \begin{minipage}{0.925\marginparwidth}
      \textbf{Key terms} \\
      \vspace{1pc} \textbf{Mechanismal explanations:} explanations of AI model behaviour which represent facts about the underlying mechanisms of prediction, such as the model structure, training data, or model weights. They are generated from introspection of the model, its training data, and its inference process. Examples include LIME and SHAP. \\
      \vspace{1pc} \textbf{Exoplanations:} statements which appear to be explanations of AI output, but are not (and cannot be) a grounded reflection of the mechanism that generated that output. This is what language models produce when asked to ``explain'' themselves. \\
      % \vspace{1pc} \textbf{Images \& Figures:} Practically anything
      % can be put in the margin if it fits. Use the
      % \texttt{{\textbackslash}marginparwidth} constant to set the
      % width of the figure, table, minipage, or whatever you are trying
      % to fit in this skinny space.
    \end{minipage}}\label{sec:sidebar} }

Classic examples of mechanismal explanations include LIME \cite{ribeiro2016should}, SHAP \cite{lundberg2017unified}, saliency maps \cite{zeiler2014visualizing, simonyan2013deep}, and Kulesza et al.'s visualisations of Bayes classifiers \cite{kulesza2015principles}, and Sarkar et al.'s visualisations of k-NN models \cite{sarkar2015interactive}. Mechanismal explanations are not the only kind: researchers in recent years have carefully drawn attention to aspects of AI explanation that instead pertain to the socio-technical system in which AI is embedded \cite{ehsan2021explainable, ehsan2021expanding, ehsan2023charting}.

A disconnection is now immediately visible between a classic mechanismal explanation, and what is produced when a language model is prompted to generate an explanation. The former is truly generated from, and has a concrete, grounded relation to, the actual processes and behaviour of a model. But a language model ``explanation'' has no such property. This has been previously noted \cite{bommasani2021opportunities, liao2024AI}, but the reason for the problem is treated as self-evident. I would like to expand on these observations, to explain why so-called ``self-explanations'' are considered to be ungrounded.

Let us examine what is actually happening when a language model has produced some output O, and is then prompted to give an explanation E for O. The process of generating E is simply another execution of the language model. E is a text composed through a sequence of next-token predictions, stochastically optimised to satisfy the query. E is not the result of an introspective reflection on the algorithmic process that was followed to generate O. A true mechanismal explanation would invoke, for example, some reference to the actual training data, model parameters, or activations, that were involved in the production of O. But if this meta-information about the prediction process is not accessible to the model to draw upon in generating E, it is theoretically impossible that E could reflect it, accurately or otherwise.

The situation is no different if E and O are requested in a single prompt, e.g., \emph{``What is the capital of France? Explain your answer.''}, as opposed to two separate prompts or conversational turns, the first asking the question and the second asking for the explanation. In the single-turn case, and in multi-turn systems where previous responses are included in the query context, it is true that the generation of O is affected by the presence of an E-request in the query context, and vice versa. For example, the language model may well produce a more coherent and well-justified output if it can simultaneously attend to a fragment of language in which an explanation is requested. However, the notional E portion of the response still lacks a \emph{mechanismal} grounding.

Statements of type E, then, are not explanations, at least not in the sense that the word is most commonly used in explainable AI research, and, we shall see, not even in the sense that users colloquially expect from these systems. They do not hold the epistemically privileged status over statements of the type O that they claim or that people expect. In fact, they are outputs like any other. E-type statements could be described as justifications, or ``post-hoc rationalisations'' \cite{rajani2019explain}, but even these terms imply a greater degree of reflexivity and introspection than is warranted. They are simulacra of justification, or of rationalisation; samples from the space of texts with the shape of justifications.

Let us instead call them \emph{exoplanations}. This term retains all the connotations of explanations (they may or may not be correct, they carry the appearance of insight, they often appeal to cause, logic, or authority), but explicitly captures the fact that they are \emph{exo}genous to, outside of, the output they explain. They inhabit the same plane of reasoning as their object; they cannot look any further beneath the object than the object itself can.

% \todo[inline]{Previous work on whether large language models can explain themselves goes here.}

Despite state-of-the-art performance on reasoning tasks \cite{rajani2019explain,wei2022chain,zhao2023self}, and one study that reported feature attribution explanations with performance comparable to LIME \cite{huang2023can}, recent work has delivered significant evidence that language models consistently fail to accurately explain their own output, and can even systematically misrepresent the true reason for a model's prediction \cite{bubeck2023sparks,madsen2024selfexplanations,sherburn2024language,turpin2024language}. In other words, at present, when the explanation sought requires introspection into the generation process, exoplanations just don't work. Large language models cannot explain themselves.

This does not mean that exoplanations are not useful; on the contrary, when presented appropriately, they can be an important and powerful tool in the designer's toolkit for creating useful and trustworthy experiences. Before we discuss those, let us turn our attention briefly to why it is important to make the distinction between exoplanations and explanations, beyond academic pedantry.

% Before we discuss those, let us briefly discuss why it is important to make the distinction between exoplanations and explanations.

% Before we discuss those, let us turn our attention briefly to why it is important to make the distinction between exoplanations and explanations, beyond academic pedantry.

% \clearpage
\section{Societal harms of exoplanations}
    
The story of the New York lawyers who submitted a legal brief including case citations generated by ChatGPT, but which turned out to be non-existent, is now well-known \cite{Merken_2023}. 

A less well-known aspect of this episode is that the infelicitous lawyers did attempt to verify that the cases were real... by asking ChatGPT, which confidently exoplained that the cases were real: \emph{``[The lawyer] asked the AI tool whether [a case it generated] is a real case. ChatGPT answered that it "is a real case" and "can be found on legal research databases such as Westlaw and LexisNexis." When asked if the other cases provided by ChatGPT are fake, it answered, "No, the other cases I provided are real and can be found in reputable legal databases such as LexisNexis and Westlaw."''} \cite{Brodkin_2023}

It has often been noted that language model hallucinations are particularly dangerous because of the bold confidence with which the model makes its assertions. The same is true of exoplanations. Because it is so easy to prompt a language model to produce an exoplanation, which is reported with bold confidence, the user can be forgiven for thinking that exoplanations are mechanismal explanations, whereas in fact they are not. This can lead to the very obvious problems such as the example above. As the firm stated in response to the judgment that the lawyers had acted in bad faith, \emph{``We made a good faith mistake in failing to believe that a piece of technology could be making up cases out of whole cloth''} \cite{Merken_2023}. 

As designers we must ask ourselves: in whom (or what) was this ``good faith'' placed, and why? If a false statement presented with bold confidence is dangerous, a false statement presented and exoplained with bold confidence is doubly so. Research in social psychology has shown that additional information can increase persuasiveness, even if it is irrelevant to the request \cite{langer1978mindlessness}. Users are easily influenced and can place their trust in meaningless explanations \cite{eiband2019impact}, and can over-trust interpretability aids \cite{kaur2020interpreting}. Allowing a system to present exoplanations with the veneer of explanations, in a situation where the user expects an explanation, should therefore be considered a dark pattern \cite{brignull-2023}.

% There are also other harms of such exoplanations. They are: ...?

The illusion of explanation perpetuated by exoplanations poses a threat to decision-making processes, in everyday knowledge work as well as in high-stakes environments such as legal or medical contexts. Reliance on exoplanations may diminish users' critical thinking and decision-making abilities.

Instead of engaging in introspection or evaluating the logic and evidence behind the model's output, users may accept exoplanations at face value. And why shouldn't they? Computers are tools, and tools are not viewed as being adversarial to the activity they facilitate. It does not seem to be a productive avenue for interaction design to attempt to erase the cultural, inertial tendency to trust computers as computationally correct machines, even if that tendency is wildly misplaced in language models.

% This can lead to a decreased sense of agency and an overreliance on AI systems for decision-making. Users can become passive recipients of information, accepting exoplanations without questioning their validity or seeking additional sources of verification.

% Another societal harm is the potential for exacerbating existing biases and inequalities. Language models trained on biased datasets may produce exoplanations that perpetuate and reinforce societal prejudices.

Exoplanations can also impair user trust and confidence in AI systems in the long term. As exoplanations are revealed to not, in fact, have their putative explanatory power, this can erode trust, and undermine any legitimate credibility that AI systems might have.

% but also diminishes user agency and autonomy in decision-making.

\marginpar{%
\RaggedRight{} 
  \vspace{-340pt} \fbox{%
    \begin{minipage}{0.925\marginparwidth}
      \textbf{Harms of exoplanations} \\
      \vspace{1pc} \textbf{False confidence:} bold exoplanations of hallucinated statements can give users false confidence in those statements, with dangerous consequences. \\
      \vspace{1pc} \textbf{Diminished critical thinking:} instead of engaging in introspection or evaluating the logic and evidence behind the model's output, users may accept exoplanations at face value. \\
      \vspace{1pc} \textbf{Erosion of trust:} when users discover that exoplanations do not accurately explain language model behaviour, this can undermine the credibility of AI systems.
    \end{minipage}}\label{sec:sidebar} }
    
% \clearpage
\section{Recontextualising ex(o)planations}

% \marginpar{%
%   \vspace{-45pt} \fbox{%
%     \begin{minipage}{0.925\marginparwidth}
%       \textbf{What to do about it} \\
%       \vspace{1pc} \textbf{Preparation:} Do not change the margin
%       dimensions and do not flow the margin text to the
%       next page. \\
%       \vspace{1pc} \textbf{Materials:} The margin box must not intrude
%       or overflow into the header or the footer, or the gutter space
%       between the margin paragraph and the main left column. The text
%       in this text box should remain the same size as the body
%       text. Use the \texttt{{\textbackslash}vspace{}} command to set
%       the margin
%       note's position. \\
%       \vspace{1pc} \textbf{Images \& Figures:} Practically anything
%       can be put in the margin if it fits. Use the
%       \texttt{{\textbackslash}marginparwidth} constant to set the
%       width of the figure, table, minipage, or whatever you are trying
%       to fit in this skinny space.
%     \end{minipage}}\label{sec:sidebar} }
    
This is clearly a case that calls for a social construction of explainability, which should \emph{``start with ``who'' the relevant stakeholders are, their explainability needs, and justify how
a particular conception of explainability satisfies the shared goals
of the relevant social group''} \cite{ehsan2022social}.

% \todo[inline]{Previous work on mechanismal explanations for LLMs, attention-based, etc. can go here.}

It is not that mechanismal explanations for language models are lacking. Despite significant challenges \cite{sarkar2022explainable}, numerous techniques have been developed to explain feature attribution, neuron activation, model attention, etc. \cite{zhao2024explainability,liao2024AI}.

However, mechanismal explanations are not the aim in and of themselves; the important aspect of the user experience that explanations need to fulfil is \emph{decision support} \cite{sarkar2015interactive, sarkar2016phd, liao2024AI, fok2024search}. Is the output correct? If it isn't, what do I need to do to fix it? Can I trust this? For example, in an AI system that generates spreadsheet formulas from natural language queries, it is by far more important and consequential for the user experience to explain the generated \emph{formula}, what it does and how it works, rather than the \emph{mechanism} of the language model that produced it. Mechanismal explanations may generate confusion and information overload in such a context \cite{kulesza2013too, liao2024AI}. 

As Miller \cite{miller2019explanation} and Sarkar \cite{sarkar2022explainable} have noted, human-human explanations are generally not mechanismal, in the sense that human-generated explanations of human behaviour rarely invoke low-level psychological or neurological phenomena, yet they are still generally successful at fulfilling the needs of everyday communication. Effective explanations can be contrastive, counterfactual, and justificatory, with respect to some intended state of affairs; these have nothing to do with the causal mechanisms underlying behaviour.

Parts of the decision support problem can be addressed though an approach termed ``co-audit'' \cite{gordon2023co}: tools to help check AI-generated content. An example of this would be the ``grounded utterances'' generated through a separate and deterministic mechanism  to explain the model output \cite{liu2023wants}. Another technique, employed by Microsoft Copilot (formerly Bing Chat) is to cite references to its Web sources that can be followed and verified. These are true explanations: they rely on mechanisms and authorities separate from the model itself and with an epistemically privileged view over the output generation process.

But exoplanations themselves can also be useful. Without needing to introspect the model, they can generate statements which help the \emph{user} rationalise, justify, and evaluate. They can generate text that prompts the \emph{user} to reflect on the output and their intents. Exoplanations can thus promote critical thinking about interactions with generative AI \cite{sarkar2024cacm}.

I propose a simple design implication that can be applied immediately: the introduction of guardrails and interface warnings against exoplanations. Commercial systems such as ChatGPT already abound with guardrails against content deemed inappropriate by the system designers, such as violent or sexual content, and numerous disclaimers against hallucinations, to the effect of \emph{``AI generated content may be incorrect.''} To these considerations, I suggest adding guardrails against exoplanations masquerading as explanations, and contextualising them to allow their true and appropriate utility to shine. 

For example, if the user asks the system to explain its output, it could produce a disclaimer of the following type: \emph{``You asked for an explanation, but as a language model, I am incapable of explaining my own behaviour.''} It might then follow this with \emph{``However, I can provide examples of how to justify, rationalise, or evaluate my previous response. Here are example arguments for and against it. This is not an explanation of my previous response.''} Together, such a disclaimer followed by an exoplanation could help defuse the worst dangers and infuse some critical thought.

% \todo[inline]{Why we believe this kind of provocation can work.}
There is reason to believe that such simple interventions can have a meaningful effect. The presence of metacognitive guiding questions, such as ``what do I understand from the text so far?'' significantly improves reading comprehension \cite{salomon1988ai}. Framing explanations as questions improves human logical discernment \cite{danry2023don}. When technology sparks conflict in discussions, it improves critical thinking \cite{lee2023fostering}. Users are influenced by the language of conversational systems and can change their instructional vocabulary and grammar after just a single exposure to system output \cite{liu2023wants}. The very same forces that influence and nudge users into trusting false explanations can be marshalled for their benefit instead.

Going forward, as more \emph{true} explanation mechanisms are developed: co-audit tools, grounded utterances, citations, etc., such disclaimers may be replaced with more concrete decision-support mechanisms. However, the utility of exoplanations as critical thinking support will remain. The key will be in helping the user develop safe and effective behaviours and mental models of trust around the different sources of evaluation and reflection available.

\section{Acknowledgements}
Thanks to my reviewers for their time and feedback.

\balance{} 

\clearpage
\bibliographystyle{SIGCHI-Reference-Format}
\bibliography{references}


\begin{thebibliography}{00}

%%% the Web address.

\ifx \showCODEN    \undefined \def \showCODEN     #1{\unskip}     \fi
\ifx \showDOI      \undefined \def \showDOI       #1{{\tt DOI:}\penalty0{#1}\ } \fi
\ifx \showISBNx    \undefined \def \showISBNx     #1{\unskip}     \fi
\ifx \showISBNxiii \undefined \def \showISBNxiii  #1{\unskip}     \fi
\ifx \showISSN     \undefined \def \showISSN      #1{\unskip}     \fi
\ifx \showLCCN     \undefined \def \showLCCN      #1{\unskip}     \fi
\ifx \shownote     \undefined \def \shownote      #1{#1}          \fi
\ifx \showarticletitle \undefined \def \showarticletitle #1{#1}   \fi
\ifx \showURL      \undefined \def \showURL       #1{#1}          \fi

\bibitem{bommasani2021opportunities}
{Rishi Bommasani}, {Drew~A Hudson}, {Ehsan Adeli}, {Russ Altman}, {Simran Arora}, {Sydney von Arx}, {Michael~S Bernstein}, {Jeannette Bohg}, {Antoine Bosselut}, {Emma Brunskill}, {and} {others}. 2021.
\newblock \showarticletitle{On the opportunities and risks of foundation models}.
\newblock {\em arXiv preprint arXiv:2108.07258\/} (2021).
\newblock


\bibitem{brignull-2023}
{H Brignull}, {M Leiser}, {C Santos}, {and} {K Doshi}. 2023.
\newblock {Deceptive patterns – user interfaces designed to trick you}.
\newblock   (4 2023).
\newblock
\showURL{%
\url{https://www.deceptive.design/}}


\bibitem{Brodkin_2023}
{Jon Brodkin}. 2023.
\newblock Lawyer cited 6 fake cases made up by CHATGPT; judge calls it “unprecedented”.
\newblock   (Jun 2023).
\newblock
\showURL{%
\url{https://arstechnica.com/tech-policy/2023/05/lawyer-cited-6-fake-cases-made-up-by-chatgpt-judge-calls-it-unprecedented/}}


\bibitem{bubeck2023sparks}
{S{\'e}bastien Bubeck}, {Varun Chandrasekaran}, {Ronen Eldan}, {Johannes Gehrke}, {Eric Horvitz}, {Ece Kamar}, {Peter Lee}, {Yin~Tat Lee}, {Yuanzhi Li}, {Scott Lundberg}, {and} {others}. 2023.
\newblock \showarticletitle{Sparks of artificial general intelligence: Early experiments with gpt-4}.
\newblock {\em arXiv preprint arXiv:2303.12712\/} (2023).
\newblock


\bibitem{danry2023don}
{Valdemar Danry}, {Pat Pataranutaporn}, {Yaoli Mao}, {and} {Pattie Maes}. 2023.
\newblock \showarticletitle{Don’t Just Tell Me, Ask Me: AI Systems that Intelligently Frame Explanations as Questions Improve Human Logical Discernment Accuracy over Causal AI explanations}. In {\em Proceedings of the 2023 CHI Conference on Human Factors in Computing Systems}. 1--13.
\newblock


\bibitem{ehsan2021expanding}
{Upol Ehsan}, {Q~Vera Liao}, {Michael Muller}, {Mark~O Riedl}, {and} {Justin~D Weisz}. 2021a.
\newblock \showarticletitle{Expanding explainability: Towards social transparency in ai systems}. In {\em Proceedings of the 2021 CHI Conference on Human Factors in Computing Systems}. 1--19.
\newblock


\bibitem{ehsan2021explainable}
{Upol Ehsan}, {Samir Passi}, {Q~Vera Liao}, {Larry Chan}, {I Lee}, {Michael Muller}, {Mark~O Riedl}, {and} {others}. 2021b.
\newblock \showarticletitle{The who in explainable ai: How ai background shapes perceptions of ai explanations}.
\newblock {\em arXiv preprint arXiv:2107.13509\/} (2021).
\newblock


\bibitem{ehsan2022social}
{Upol Ehsan} {and} {Mark~O. Riedl}. 2022.
\newblock Social Construction of XAI: Do We Need One Definition to Rule Them All?
\newblock   (2022).
\newblock


\bibitem{ehsan2023charting}
{Upol Ehsan}, {Koustuv Saha}, {Munmun De~Choudhury}, {and} {Mark~O Riedl}. 2023.
\newblock \showarticletitle{Charting the Sociotechnical Gap in Explainable AI: A Framework to Address the Gap in XAI}.
\newblock {\em Proceedings of the ACM on Human-Computer Interaction\/} {7}, CSCW1 (2023), 1--32.
\newblock


\bibitem{eiband2019impact}
{Malin Eiband}, {Daniel Buschek}, {Alexander Kremer}, {and} {Heinrich Hussmann}. 2019.
\newblock \showarticletitle{The Impact of Placebic Explanations on Trust in Intelligent Systems}. In {\em Extended Abstracts of the 2019 CHI Conference on Human Factors in Computing Systems} {\em (CHI EA '19)}. Association for Computing Machinery, New York, NY, USA, 1–6.
\newblock
\showISBNx{9781450359719}
\showDOI{%
\url{http://dx.doi.org/10.1145/3290607.3312787}}


\bibitem{fok2024search}
{Raymond Fok} {and} {Daniel~S. Weld}. 2024.
\newblock In Search of Verifiability: Explanations Rarely Enable Complementary Performance in AI-Advised Decision Making.
\newblock   (2024).
\newblock


\bibitem{gordon2023co}
{Andrew~D Gordon}, {Carina Negreanu}, {Jos{\'e} Cambronero}, {Rasika Chakravarthy}, {Ian Drosos}, {Hao Fang}, {Bhaskar Mitra}, {Hannah Richardson}, {Advait Sarkar}, {Stephanie Simmons}, {and} {others}. 2023.
\newblock \showarticletitle{Co-audit: tools to help humans double-check AI-generated content}.
\newblock {\em arXiv preprint arXiv:2310.01297\/} (2023).
\newblock


\bibitem{huang2023can}
{Shiyuan Huang}, {Siddarth Mamidanna}, {Shreedhar Jangam}, {Yilun Zhou}, {and} {Leilani~H Gilpin}. 2023.
\newblock \showarticletitle{Can large language models explain themselves? a study of llm-generated self-explanations}.
\newblock {\em arXiv preprint arXiv:2310.11207\/} (2023).
\newblock


\bibitem{kaur2020interpreting}
{Harmanpreet Kaur}, {Harsha Nori}, {Samuel Jenkins}, {Rich Caruana}, {Hanna Wallach}, {and} {Jennifer Wortman~Vaughan}. 2020.
\newblock \showarticletitle{Interpreting Interpretability: Understanding Data Scientists' Use of Interpretability Tools for Machine Learning}. In {\em Proceedings of the 2020 CHI Conference on Human Factors in Computing Systems} {\em (CHI '20)}. Association for Computing Machinery, New York, NY, USA, 1–14.
\newblock
\showISBNx{9781450367080}
\showDOI{%
\url{http://dx.doi.org/10.1145/3313831.3376219}}


\bibitem{kulesza2015principles}
{Todd Kulesza}, {Margaret Burnett}, {Weng-Keen Wong}, {and} {Simone Stumpf}. 2015.
\newblock \showarticletitle{Principles of explanatory debugging to personalize interactive machine learning}. In {\em Proceedings of the 20th international conference on intelligent user interfaces}. 126--137.
\newblock


\bibitem{kulesza2013too}
{Todd Kulesza}, {Simone Stumpf}, {Margaret Burnett}, {Sherry Yang}, {Irwin Kwan}, {and} {Weng-Keen Wong}. 2013.
\newblock \showarticletitle{Too much, too little, or just right? Ways explanations impact end users' mental models}. In {\em 2013 IEEE Symposium on visual languages and human centric computing}. IEEE, 3--10.
\newblock


\bibitem{langer1978mindlessness}
{Ellen~J Langer}, {Arthur Blank}, {and} {Benzion Chanowitz}. 1978.
\newblock \showarticletitle{The mindlessness of ostensibly thoughtful action: The role of" placebic" information in interpersonal interaction.}
\newblock {\em Journal of personality and social psychology\/} {36}, 6 (1978), 635.
\newblock


\bibitem{lee2023fostering}
{Sunok Lee}, {Dasom Choi}, {Minha Lee}, {Jonghak Choi}, {and} {Sangsu Lee}. 2023.
\newblock \showarticletitle{Fostering Youth’s Critical Thinking Competency About AI through Exhibition}. In {\em Proceedings of the 2023 CHI Conference on Human Factors in Computing Systems} {\em (CHI '23)}. Association for Computing Machinery, New York, NY, USA, Article 451, 22 pages.
\newblock
\showISBNx{9781450394215}
\showDOI{%
\url{http://dx.doi.org/10.1145/3544548.3581159}}


\bibitem{liao2024AI}
{Q.~Vera Liao} {and} {Jennifer Wortman~Vaughan}. 2024.
\newblock \showarticletitle{{AI} {Transparency} in the {Age} of {LLMs}: A {Human}-{Centered} {Research} {Roadmap}}.
\newblock {\em Harvard Data Science Review\/} (feb 29 2024).
\newblock
\newblock
\shownote{https://hdsr.mitpress.mit.edu/pub/aelql9qy.}


\bibitem{liu2023wants}
{Michael~Xieyang Liu}, {Advait Sarkar}, {Carina Negreanu}, {Benjamin Zorn}, {Jack Williams}, {Neil Toronto}, {and} {Andrew~D Gordon}. 2023.
\newblock \showarticletitle{“What It Wants Me To Say”: Bridging the Abstraction Gap Between End-User Programmers and Code-Generating Large Language Models}. In {\em Proceedings of the 2023 CHI Conference on Human Factors in Computing Systems}. 1--31.
\newblock


\bibitem{lundberg2017unified}
{Scott~M Lundberg} {and} {Su-In Lee}. 2017.
\newblock \showarticletitle{A unified approach to interpreting model predictions}.
\newblock {\em Advances in neural information processing systems\/}  {30} (2017).
\newblock


\bibitem{madsen2024selfexplanations}
{Andreas Madsen}, {Sarath Chandar}, {and} {Siva Reddy}. 2024.
\newblock Are self-explanations from Large Language Models faithful?
\newblock   (2024).
\newblock


\bibitem{Merken_2023}
{Sara Merken}. 2023.
\newblock New York lawyers sanctioned for using fake ChatGPT cases in legal brief.
\newblock   (Jun 2023).
\newblock
\showURL{%
\url{https://www.reuters.com/legal/new-york-lawyers-sanctioned-using-fake-chatgpt-cases-legal-brief-2023-06-22/}}


\bibitem{miller2019explanation}
{Tim Miller}. 2019.
\newblock \showarticletitle{Explanation in artificial intelligence: Insights from the social sciences}.
\newblock {\em Artificial intelligence\/}  {267} (2019), 1--38.
\newblock


\bibitem{rajani2019explain}
{Nazneen~Fatema Rajani}, {Bryan McCann}, {Caiming Xiong}, {and} {Richard Socher}. 2019.
\newblock \showarticletitle{Explain yourself! leveraging language models for commonsense reasoning}.
\newblock {\em arXiv preprint arXiv:1906.02361\/} (2019).
\newblock


\bibitem{ribeiro2016should}
{Marco~Tulio Ribeiro}, {Sameer Singh}, {and} {Carlos Guestrin}. 2016.
\newblock \showarticletitle{"Why should I trust you?" Explaining the predictions of any classifier}. In {\em Proceedings of the 22nd ACM SIGKDD international conference on knowledge discovery and data mining}. 1135--1144.
\newblock


\bibitem{salomon1988ai}
{Gavriel Salomon}. 1988.
\newblock \showarticletitle{AI in reverse: Computer tools that turn cognitive}.
\newblock {\em Journal of educational computing research\/} {4}, 2 (1988), 123--139.
\newblock


\bibitem{sarkar2016phd}
{Advait Sarkar}. 2016.
\newblock {\em {Interactive analytical modelling}}.
\newblock {T}echnical {R}eport UCAM-CL-TR-920. University of Cambridge, Computer Laboratory.
\newblock
\showDOI{%
\url{http://dx.doi.org/10.48456/tr-920}}


\bibitem{sarkar2022explainable}
{Advait Sarkar}. 2022.
\newblock \showarticletitle{{Is explainable AI a race against model complexity?}}. In {\em {Workshop on Transparency and Explanations in Smart Systems (TeXSS), in conjunction with ACM Intelligent User Interfaces (IUI 2022)}} {\em (CEUR Workshop Proceedings)}. 192--199.
\newblock
\showURL{%
\url{http://ceur-ws.org/Vol-3124/paper22.pdf}}


\bibitem{sarkar2024cacm}
{Advait Sarkar}. 2024.
\newblock \showarticletitle{AI Should Challenge, Not Obey}.
\newblock {\em Communications of the ACM (in press)\/} (2024).
\newblock


\bibitem{sarkar2015interactive}
{Advait Sarkar}, {Mateja Jamnik}, {Alan~F. Blackwell}, {and} {Martin Spott}. 2015.
\newblock \showarticletitle{Interactive visual machine learning in spreadsheets}. In {\em 2015 IEEE Symposium on Visual Languages and Human-Centric Computing (VL/HCC)}. 159--163.
\newblock
\showDOI{%
\url{http://dx.doi.org/10.1109/VLHCC.2015.7357211}}


\bibitem{sherburn2024language}
{Dane Sherburn}, {Bilal Chughtai}, {and} {Owain Evans}. 2024.
\newblock Language Models Struggle to Explain Themselves.
\newblock   (2024).
\newblock
\showURL{%
\url{https://openreview.net/forum?id=o6eUNPBAEc}}


\bibitem{simonyan2013deep}
{Karen Simonyan}, {Andrea Vedaldi}, {and} {Andrew Zisserman}. 2013.
\newblock \showarticletitle{Deep inside convolutional networks: Visualising image classification models and saliency maps}.
\newblock {\em arXiv preprint arXiv:1312.6034\/} (2013).
\newblock


\bibitem{turpin2024language}
{Miles Turpin}, {Julian Michael}, {Ethan Perez}, {and} {Samuel Bowman}. 2024.
\newblock \showarticletitle{Language models don't always say what they think: unfaithful explanations in chain-of-thought prompting}.
\newblock {\em Advances in Neural Information Processing Systems\/}  {36} (2024).
\newblock


\bibitem{wei2022chain}
{Jason Wei}, {Xuezhi Wang}, {Dale Schuurmans}, {Maarten Bosma}, {Fei Xia}, {Ed Chi}, {Quoc~V Le}, {Denny Zhou}, {and} {others}. 2022.
\newblock \showarticletitle{Chain-of-thought prompting elicits reasoning in large language models}.
\newblock {\em Advances in neural information processing systems\/}  {35} (2022), 24824--24837.
\newblock


\bibitem{zeiler2014visualizing}
{Matthew~D Zeiler} {and} {Rob Fergus}. 2014.
\newblock \showarticletitle{Visualizing and understanding convolutional networks}. In {\em Computer Vision--ECCV 2014: 13th European Conference, Zurich, Switzerland, September 6-12, 2014, Proceedings, Part I 13}. Springer, 818--833.
\newblock


\bibitem{zhao2024explainability}
{Haiyan Zhao}, {Hanjie Chen}, {Fan Yang}, {Ninghao Liu}, {Huiqi Deng}, {Hengyi Cai}, {Shuaiqiang Wang}, {Dawei Yin}, {and} {Mengnan Du}. 2024.
\newblock \showarticletitle{Explainability for large language models: A survey}.
\newblock {\em ACM Transactions on Intelligent Systems and Technology\/} {15}, 2 (2024), 1--38.
\newblock


\bibitem{zhao2023self}
{Jiachen Zhao}, {Zonghai Yao}, {Zhichao Yang}, {and} {Hong Yu}. 2023.
\newblock \showarticletitle{SELF-EXPLAIN: Teaching Large Language Models to Reason Complex Questions by Themselves}.
\newblock {\em arXiv preprint arXiv:2311.06985\/} (2023).
\newblock


\end{thebibliography}

\end{document}